\documentclass[aps,twocolumn,prd,showpacs,nofootinbib,floatfix]{revtex4-1}
\usepackage{latexsym}
\usepackage{psfrag}
\usepackage{graphicx,amsmath,amsfonts,amssymb,dcolumn,epsfig,bm,float}
\usepackage{epsfig}%
\usepackage{color}
\usepackage{graphics}
\usepackage{epstopdf}
\usepackage[margin=0.5in]{geometry}
\usepackage[hang,small,bf]{caption}
\usepackage{rotating}
\usepackage{subfigure}
\usepackage{silence}
\usepackage{appendix}
\usepackage{tabularx}
\usepackage{multirow}
\usepackage{pifont}
\usepackage{lipsum}
\def\beq{\begin{equation}}
\def\eeq{\end{equation}}
\def\bea{\begin{eqnarray}}
\def\eea{\end{eqnarray}}

\begin{document}

\title{Stability of charged particles inside a Paul trap with spontaneous localization dynamics}

\author{Sayantani Bera} \email{sayantani.bera@tifr.res.in}

\author{Shreya Banerjee} \email{shreya.banerjee@tifr.res.in}


\affiliation{ Tata Institute of Fundamental Research \\
Homi Bhabha Road, Mumbai 400005, India}
\date{\today}
\begin{abstract}
\noindent Paul traps are ion traps that are widely used in spectroscopic experiments to confine and stabilize a charged particle within a small region using oscillating electric fields. The dynamics of the particle inside a Paul trap is described by Mathieu equations. It has been proposed that such traps can be used to detect the effects produced by spontaneous collapse of the associated wavefunction, as described by the model of CSL (Continuous Spontaneous Localization). This model is a non-linear, stochastic and non-relativistic modification to the Schr\"{o}dinger equation which predicts an additional random motion of particles other than environmental effects. In this paper, we discuss the possibility that such a random motion can throw a particle out of its stable configuration within the Paul trap. We study the changes in the stability diagram of a Paul trap in the presence of CSL. We also constrain the CSL parameter space by assuming the fact that the stability diagram is not significantly altered. The bounds thus obtained are weaker than those coming from X-ray emission from Ge slab. \\

\end{abstract}

\maketitle


\section{Introduction}
The enormous success of quantum theory has led many physicists believe that quantum theory is the ultimate theory and all other theories, including gravity, should be described in the language of quantum theory. While this is true to some extent, one can safely say that the picture is not complete yet. There are many glitches in quantum theory that remain unanswered, and mostly remain ignored by the community. The theory does not explain the collapse of the wavefunction during a measurement process. Neither does it allow a natural transition between quantum to classical phenomena. The superposition principle is violated during a measurement and it is simply taken as an outcome, not as a consequence of the theory. The theory also fails to address the origin of probabilistic outcomes which obey Born probability rule. The crucial role played by the measuring apparatus which is ``classical" is also very discomforting since, in an ideal picture where everything is described by quantum theory, even the measuring apparatus should follow the rules of quantum mechanics. These set of problems together are known as the ``quantum measurement problem" \cite{Bell}. One way to address these set of questions is to completely reinterpret quantum theory such that known observations are kept intact. Bohmian mechanics is one such approach \cite{Bohm}. The theory considers the coexistence of particles and wavefunctions, and thus there is no concept of collapse in the theory. The decoherence compensated by many worlds interpretation is another approach where quantum theory is treated as sacred \cite{Zurek,Everett}. The collapse of the outcomes are attributed to observers in different branches of the universe. The third approach is to extend quantum theory beyond linear regime in such a way that the collapse is encoded in the dynamics of the theory rather than depending upon measurements taking place. Theories like GRW (Ghirardi-Rimini-Weber), QMUPL (Quantum Mechanics with Universal Position Localization) and CSL (Continuous Spontaneous Localization) fall in this category \cite{Ghirardi:86,Ghirardi}. These theories are nonlinear and stochastic in nature and can naturally explain the Born probability rule. The equations are formulated such that quantum theory and classical theory are recovered in appropriate limits. For a general review on such theories, one can refer to \cite{reviewbassi,review}.

CSL theory is the most advanced form of its previous versions. A lot of interests have grown in recent times in testing CSL theory in laboratory experiments via observation of the breakdown of superposition principle in the macroscopic limit. Experiments on molecular interference are pushing the limits in the mass spectrum to observe the breakdown of quantum theory \cite{Arndt}. Other avenues to test CSL are also being explored, for example, tests of CSL-induced spectral line broadening \cite{Hendrik}, excess heat measurement via temperature increase \cite{thermo} and bounds deduced from heating of an atomic Bose-Einstein Condensate \cite{BEC}. Such proposals rely upon the ``side-effects" of CSL theory, which allow a small violation of energy-momentum conservation. CSL assumes the presence of a universal noise that couples with the matter and imparts an excess energy to the system. As a result, a particle affected by CSL can undergo a Brownian motion or a slight increase in its temperature. These effects should be detectable in an experiment given other sources of noise are minimized. Recent proposals on the detection of heating effect and resulting random walk have been discussed in \cite{Bhawna,Bedingham,Adler}. In a very recent experiment reported in \cite{Vinante}, the authors claim to have observed an extra noise source. If this could be connected to CSL, then that would be a significant step towards understanding the  fundamental theory. Various bounds on the CSL parameters have already restricted the parameter space to a large extent, and it is only a matter of time when the theory would be validated/ ruled out by experiments. Given the surge of interest in testing CSL, that does not seem far-fetched. 

In this letter, we rather try to point out basic readjustments in the apparatus that need to be done in order to test CSL and similar theories. Most of the proposals of detection rely on a trapping mechanism where particles undergoing CSL are trapped optically or via oscillating electric fields. The most widely used ion traps in this context are Paul traps that use transverse electric fields oscillating in radio-frequency and producing a stable field configuration to hold the charged particle. In this work, we consider scenarios where the stability of the particle might get disturbed due to additional random motion caused by CSL. It can even drive the particle out of its stability zone. Thus to perform such experiments, one must first consider the backreaction of CSL effect on the particle stability inside the trap and readjust the trap parameters such that throughout the experiment the particle is confined inside the trap. We solve the modified trajectory of the particle with CSL noise and plot the resulting stability zone. We also draw an exclusion plot of the CSL parameters for a given trap configuration.
\section{Trajectory of an ion in Paul trap}
Paul traps involve time dependent cross electric fields to restrict the motion of a charged particle and hence trapping the ion inside a small region. The most widely used Paul traps are quadrupolar Paul traps where the field applied is of quadrupolar nature. The potential depends upon a time varying voltage applied to the electrodes on top of a static dc voltage. The restoring force on the ion increases linearly with its distance from the centre, thus effectively confining the ion at the centre of the trap. Here we consider the 2D quadrupolar Paul trap which was considered in \cite{Bedingham}. Such ion traps confine ions in two direction while the third direction is used to eject ions inside the trap. Figure below depicts the basic structure of a 2D Paul trap. 

\begin{figure}[H]
\centering
  \includegraphics[width=8cm,height=7cm] {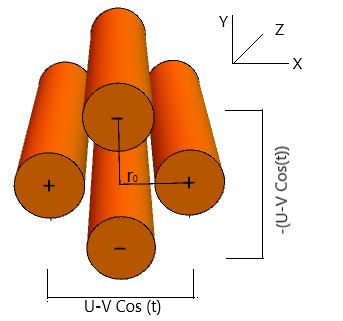}
\caption[Optional caption for list of figures]{\small Schematic diagram of a 2-D quadrupolar Paul trap. An oscillating electric field is applied between each pair of electrodes. In the diagram, $X-Y$ is the confining plane, and particles are inserted into the trap along the $Z$-direction.} 
\end{figure}

There are four electrodes, each pair of opposite rods having the same polarity. This structure creates a quadrupolar field in the $x-y$ plane. The distance from the centre of the trap to the surface of the electrodes is $r_0$. A potential $U-V \cos(\Omega t)$ is applied to the x-electrodes, while $-(U-V \cos(\Omega t))$ is applied to the y-electrodes. $U$ is the dc voltage and $V$ denotes the zero-to-peak amplitude of an oscillating radio-frequency voltage having frequency $\Omega$. It can be shown that the equation of motion of an ion with charge $Q$ inside the trap can be written as,

\begin{equation}
m\left(\frac{d^2 x}{dt^2}\right) + \left(\frac{2QU}{r_0^2}- \frac{2QV\cos(\Omega t)}{r_0^2}\right)x = 0
\label{Mathieu1}
\end{equation}
A similar equation is valid for the y-direction.\\
This equation has the same form as the Mathieu equation which was obtained to describe the motion of a vibrating elliptical drumhead \cite{Mathieu}. The equation has the following form,
\begin{equation}
\frac{d^2 u}{d\xi^2} + \left(a_u + 2q_u\cos(2\xi)\right)u = 0
\label{Mathieu2}
\end{equation}
where $\xi = \Omega t/2$ is dimensionless parameter. $a_u$ and $q_u$ are the dimensionless parameters called the stability parameters. The solution of the Mathieu equation can be described in terms of stability zones in the $a_u-q_u$ parameter space. \\

Using the transformation $\xi = \Omega t/2$, Eq. \eqref{Mathieu1} can be written exactly in the form of the Mathieu equation \eqref{Mathieu2} with $a_x= 8 QU/m\Omega^2 r_0^2$ and $q_x=-4 QV/m\Omega^2 r_0^2$. For the equation in y-direction, we get $a_y=-a_x$ and $q_y=-q_x$.

The general solution can be written as 
\begin{equation}
u(\xi)= A e^{\mu\xi}\sum_{n=-\infty}^\infty C_{2n}e^{2in\xi} + B e^{-\mu\xi}\sum_{n=-\infty}^\infty C_{2n}e^{-2in\xi}
\end{equation}
where $A,\ B$ are the integration constants that depend upon the initial conditions, $C_{2n}$ denote the amplitude of oscillations and $\mu$ is a dimensionless parameter (Floquet index) that depends upon $a_u$ and $q_u$. This parameter is crucial to determine the stable and unstable regions of the solution. When $\mu$ is purely imaginary, i.e. $\mu=i \beta$, then those solutions are stable oscillatory solutions.\\

For the cases of our interest, we work with $|a_u| \ll |q_u| \ll 1$. The analytical solution in terms of $t$ in this regime can be obtained as (two integration constants are chosen to be same):
\begin{equation}
x(t) = A \cos(\mu \Omega t/2)\left[1+\frac{q}{2} \cos(\Omega t)\right]
\end{equation} 
where $\mu= \sqrt{a+q^2/2}$ (Dehmelt's approximation). 

\section{Continuous Spontaneous Localization and random motion}
The Continuous Spontaneous Localization or the CSL model was first proposed by Ghirardi, Pearle and Rimini in the context of modifying quantum mechanics in order to address the issues related to the ``quantum measurement problem" \cite{Ghirardi}. The proposed model is a \textit{non-relativistic} modification to quantum mechanics and a proper relativistic extension is yet to be formulated. The mass proportional version of CSL evolution equation can be written as \cite{review},

\begin{equation} 
\begin{split}
\label{eq:csl-massa}
& d\psi_t = \Big[-\frac{i}{\hbar}\hat{H}dt 
 + \frac{\sqrt{\gamma}}{m_{0}}\int d\mathbf{x} (\hat{M}(\mathbf{x}) - \langle \hat{M}(\mathbf{x}) \rangle_t)
dW_{t}(\mathbf{x})  \\ &
 ~~~~~~~~~~ - \frac{\gamma}{2m_{0}^{2}} \int d\mathbf{x}\, d\mathbf{y}\, \mathcal{G}({\bf x}-{\bf y})
(\hat{M}(\mathbf{x})-\langle \hat{M}(\mathbf{x}) \rangle_t) \times \\ & ~~~~~~~~~~~~~~~~~~~~~~~~~~~~(\hat{M}(\mathbf{y})-\langle \hat{M}(\mathbf{y}) \rangle_t)dt\Big] \psi_t  
\end{split}
\end{equation}

where $H$ represents the linear part - the standard Hamiltonian of the system, and the other two terms which are non-linear and stochastic modifications, are accountable for the collapse of the wave function. The mass $m_0$ is a reference mass i.e. the mass of a nucleon. Here $\mathcal{G}({\bf x}-{\bf y})$ is the noise two point correlation function which is taken to be Gaussian in this model. $\gamma$, a free parameter of the theory, is the positive coupling constant which sets the strength of the collapse process. $\hat{M}({\bf x})$ is the smeared mass density operator that encodes the amplification mechanism in CSL and is given as,
\begin{eqnarray}
\hat{M}({\bf x})&=& \sum_{j}m_{j}\hat{N}_{j}({\bf x}) \\
\hat{N}_{j}({\bf x})&=& \int d{\bf y}g({\bf y}-{\bf x})\hat{a}_{j}^{\dagger}({\bf y})\hat{a}_{j}({\bf y})
\end{eqnarray}
The number operator $\hat{N}_{j}$ contains $\hat{a}_{j}^{\dagger}({\bf y})$ and $\hat{a}_{j}({\bf y})$ which are, respectively, the creation and annihilation operators of a particle of type $j$ at the space point ${\bf y}$. The smearing function is defined as
\begin{equation}
g({\bf x})=\frac{1}{(\sqrt{2\pi}r_{c})^{3}}e^{-{\bf x}^{2}/2r_{c}^{2}}
\end{equation} 
The free parameter $r_c$ of the theory represents the correlation length. $W_t({\bf x})$ is an ensemble of Wiener processes representing white noise. The collapse strength $\gamma$ is related to the correlation/ collapse width $r_c$ and the collapse rate parameter $\lambda_{CSL}$ in the following way,
$$ \lambda_{CSL} = \frac{\gamma}{(4\pi r_c^2)^{3/2}}$$
$\lambda_{CSL}$ denotes the collapse rate of a single nucleon. For multi-particle systems, the rate is amplified by the total number of particles. The values of these parameters, as suggested in the original theory, are $r_c=10^{-5} cm$ and $\lambda_{CSL}= 10^{-17}s^{-1}$ (we call it $\lambda_{GRW}$). A different and a much stronger value of $\lambda_{CSL} $ was later suggested by Adler based on the observations in the latent image formation in photography, which is $ \lambda_{CSL}=10^{-8}s^{-1}$ (we call it $\lambda_{ADLER}$). Both the values satisfy known experimental observations and only future and more precise experiments can suggest which of these bounds is correct \footnote{The recent experiment reported in \cite{Vinante} is in tension with Adler's choice of parameters. In a recent paper \cite{Adler}, it has been proposed that the CSL noise should be treated as non-white with a frequency cut-off. Adler's parameter values are not ruled out if this scenario is taken into account}.

One of the major consequences of CSL dynamics that distinguishes it from quantum theory is the violation of energy-momentum conservation law. Since CSL contains a noise source that interacts with matter, we have a resulting non-conservation of energy and momentum when the system is concerned. The noise field imparts energy and momentum to the particles, causing a Brownian motion-like random motion, and corresponding increase in temperature due to this heating effect. These effects were investigated in \cite{Adler1, Pearle} and possible experiments were suggested in \cite{Bhawna,Bedingham,Goldwater,Adler}. For a sphere of radius $R$, the rms diffusion caused by CSL can be written as,
\begin{equation}
\Delta x_{CSL} = \frac{\hbar}{m_0 r_c}\left(\frac{\lambda_{CSL}f(R/r_c)}{6}\right)^{1/2} t^{3/2}
\label{CSLwalk}
\end{equation}
Notice that the displacement is not directly dependent upon the mass or density of the particle; the dependence is through the size $R$ of the system. The function $f$ is given as,
\begin{equation}
f(R/r_c)= 6 \left(\frac{r_c}{R}\right)^4 \left[1-2\frac{r_c^2}{R^2}+ \left(1+2\frac{r_c^2}{R^2}\right)e^{-R^2/r_c^2}\right]
\end{equation}
For our cases of interest, we take $R \sim r_c$ for which $f \approx 1$ \cite{Bedingham,Bhawna}.

It has been suggested in \cite{Bedingham} that the heating effect due to CSL can be detected for a particle trapped inside a Paul trap. The aim of this letter is to investigate the consequences of such heating effect in terms of the stability of the trapped ion. Since CSL induces a random motion causing a particle jitter in space, it may throw a particle out of the stable region. Thus, a charged particle, which would ideally be stable in a given Paul trap configuration, may finally become unstable and come out of the trap due to additional motions caused by CSL noise. In this article, we show how CSL affects the trajectory of an ion inside the trap and work out the resultant stability region for the suggested values of the CSL parameters.

Let us consider the motion of an ion in the x-direction. The rms displacement in the x-direction due to the CSL noise is given by \eqref{CSLwalk}. The corresponding effective force on the ion can be obtained by differentiating this twice with respect to $t$. This gives,
$$ F_{CSL}= \frac{\hbar m\sqrt{\lambda}}{m_0 r_c \sqrt{6}}\left(\frac{3}{4} t^{-1/2}-\frac{\Omega^2}{4}(a_x + 2 q_x \cos \Omega t) t^{3/2}\right) $$
The second term in the above equation is explicitly due to the extra restoring force acting on an ion as it undergoes CSL random walk, and thus depends on the trap parameters. The resultant equation of motion for an ion inside the Paul trap and acted upon by CSL is thus given by,
\begin{eqnarray}
m\ddot{x}&+&\left[a_x + 2 q_x \cos \Omega t\right]\frac{m\Omega^2}{4} x \nonumber \\ \nonumber &+&\frac{\hbar m\sqrt{\lambda}}{m_0 r_c \sqrt{6}}\left(-\frac{3}{4} t^{-1/2}+\frac{\Omega^2}{4}\left(a_x + 2 q_x \cos \Omega t\right) t^{3/2}\right)=0\\
\label{cslmathieu}
\end{eqnarray}
Using the approximation $|a_x|\ll |q_x| \ll 1$, we get an analytical solution of the above equation which looks very complicated. The form of the solution is not presented in this article to maintain brevity, since we perform a fully numerical analysis for the final results.
\section{Stability region with CSL dynamics}
As mentioned earlier, the trajectory of a charged particle inside a Paul trap can be stable or unstable based on the nature of the solution. If the solution diverges with time (i.e. when $\mu$ is real or complex), then the trajectory becomes unstable. Only for the cases when $\mu$ becomes purely imaginary, the solution is stable and periodic. Thus, the stability depends upon the trapping parameters $a$ and $q$ that define the Floquet index $\mu$.\\

In presence of CSL, things are different and the nature of the solution of the equation \eqref{cslmathieu} now also depends on the CSL parameters $\lambda_{CSL}$ and $r_c$. To see how the stability of the ion trajectory depends upon the four parameters $a$, $q$, $r_c$ and $\lambda_{CSL}$, we make use of the transfer matrix method described in \cite{ref1, ref2}. Here we briefly discuss the method. For a better understanding, readers may refer to \cite{ref1, ref2}. Transfer matrix connects the initial solution $x(0)$ and $v(0)$ with the solution at any time $t$ i.e. $x(t)$ and $v(t)$, where $x(t)$ is the displacement and $v(t)$ denotes the velocity. In matrix formalism, we can write,
\[
   \left[ {\begin{array}{c}
   x_1\\
   v_1\\
  \end{array} } \right]= M
  \left[ {\begin{array}{c}
   x_0 \\
   v_0\\
  \end{array} } \right]
\]
 where $x_0=x(0)$, $v_0=v(0)$, $x_1= x(T=2\pi/\Omega)$ (i.e. the displacement after one period of oscillation), $v_1= v(T=2\pi/\Omega)$ and $M$ is the transfer matrix. Let us consider two linearly independent solutions of the equation of motion: $u_1$ and $u_2$ . Then the general solution for $x(t)$ and $v(t)$ is given by,
 \begin{eqnarray}
 x(t)&=& A_1 u_1(t) + A_2 u_2(t) \nonumber \\
 v(t)&=& A_1 \dot{u}_1(t) + A_2 \dot{u}_2(t)\nonumber
 \end{eqnarray}
The transfer matrix $M$ can be written as \cite{ref2},
\[
   M = 
  \left[ {\begin{array}{cc}
   u_1(T) & u_2(T) \\
   \dot{u}_1(T) & \dot{u}_2(T)\\
  \end{array} } \right] \left[ {\begin{array}{cc}
   u_1(0) & u_2(0) \\
   \dot{u}_1(0) & \dot{u}_2(0)\\
  \end{array} } \right]^{-1}
\]
The transfer matrix $M$ has a property $\det(M)=1$ \cite{ref2}. For simplicity, we denote the elements of $M$ as:
\[
   M = 
   \left[ {\begin{array}{cc}
   m_{11} & m_{12} \\
   m_{21} & m_{22}\\
  \end{array} } \right]
\]
It can be shown that the solution at any time t = nT follow a recursion relation i.e,
\[
   \left[ {\begin{array}{c}
   x_n\\
   v_n\\
  \end{array} } \right]= M^n
  \left[ {\begin{array}{c}
   x_0 \\
   v_0\\
  \end{array} } \right]
\]
 Now, since $M$ is a $2 \times 2$ matrix, it has two eigen values, say $\lambda_1$ and $\lambda_2$ corresponding to two eigenvectors $m_1$ and $m_2$ respectively. The initial solutions can be expressed in terms of the eigenvectors as follows:
 \[   
   \left[ {\begin{array}{c}
   x_{0} \\
   v_{0} \\
  \end{array} } \right] = A_1 m_1 + A_2 m_2
\]
 Using the recursion relation, the solution at time $t= nT$ is obtained as,
 \[   
   \left[ {\begin{array}{c}
   x_{n} \\
   v_{n} \\
  \end{array} } \right] = A_1\lambda_1^n m_1 + A_2 \lambda_2^n m_2
\]
 We say that the solution is stable, i.e, it does not diverge after $t=nT$ if $|\lambda_1|, |\lambda_2| \leq 1$ since, otherwise $\lambda_1^n, \lambda_2^n$ becomes arbitrarily large for $n \gg 1$. This is precisely the stability condition that the solution needs to satisfy in order to represent a stable trap configuration. The values $\lambda_1,\lambda_2$ can be obtained from the following characteristic equation,
\[
   \det
   \left[ {\begin{array}{cc}
   m_{11}-\lambda & m_{12} \\
   m_{21} & m_{22}-\lambda\\
  \end{array} } \right] =0
\]
The solutions are given as
\begin{equation}
\lambda_{1,2}= s \pm i\sqrt{1-s^2}
\end{equation}
where $s= (m_{11}+m_{22})/2 = Tr(M)/2$\\
Now, for $|s| \leq 1$, we have, $|\lambda_1|,|\lambda_1| = 1$. This requires $|Tr(M)| \leq 2$. On the other hand, for $|s| > 1$, we get both $\lambda_1$ and $\lambda_2$ to be real, and one of them is always greater than $1$ making the solution diverge. Thus, the stability conditions as obtained from transfer matrix method is,
\begin{eqnarray}
|Tr(M)| &\leq & 2 \quad {\rm for ~ stable ~ solution} \nonumber \\
|Tr(M)| & >& 2 \quad {\rm for ~ unstable ~ solution} 
\end{eqnarray}

Thus, for a given set of values of $a$, $q$, $r_c$ and $\lambda_{CSL}$, the solution is said to be stable if the trace of this matrix lies between $\pm 2$. To get the stability zone in the $a-q$ parameter space, we fix $r_c$ and $\lambda_{CSL}$ to the suggested values by Ghirardi-Rimini-Weber ($r_c = 10^{-5}$ cm, $\lambda_{GRW} = 10^{-17} s^{-1}$) and Adler ($r_c = 10^{-5}$ cm, $\lambda_{ADLER} = 10^{-8} s^{-1}$) respectively, and plot the combinations $a,q$ for which $|Tr(M)| \leq 2$. The pair of points in $a-q$ plane, for which this condition is violated, correspond to unstable region. Below we show the stability zone for Adler's choices of CSL parameters. We can compare this stability plot with the usual stability region of Mathieu's equations and it is evident that there is no visible change in the stability diagram. This is good in the sense that one does not need to worry much about the alteration of ion stability in the presence of CSL. But care must be taken for certain configurations of the trap, especially for those values of $a$, $q$ which are very close to the boundary of the stable region since CSL induced random motions can make such points unstable. In fact, we do find that certain stable points move out to the unstable region if the CSL effect is made strong enough. 

\vspace{.5in}

\begin{figure}[H]
\centering
  \includegraphics[width=9cm,height=6cm] {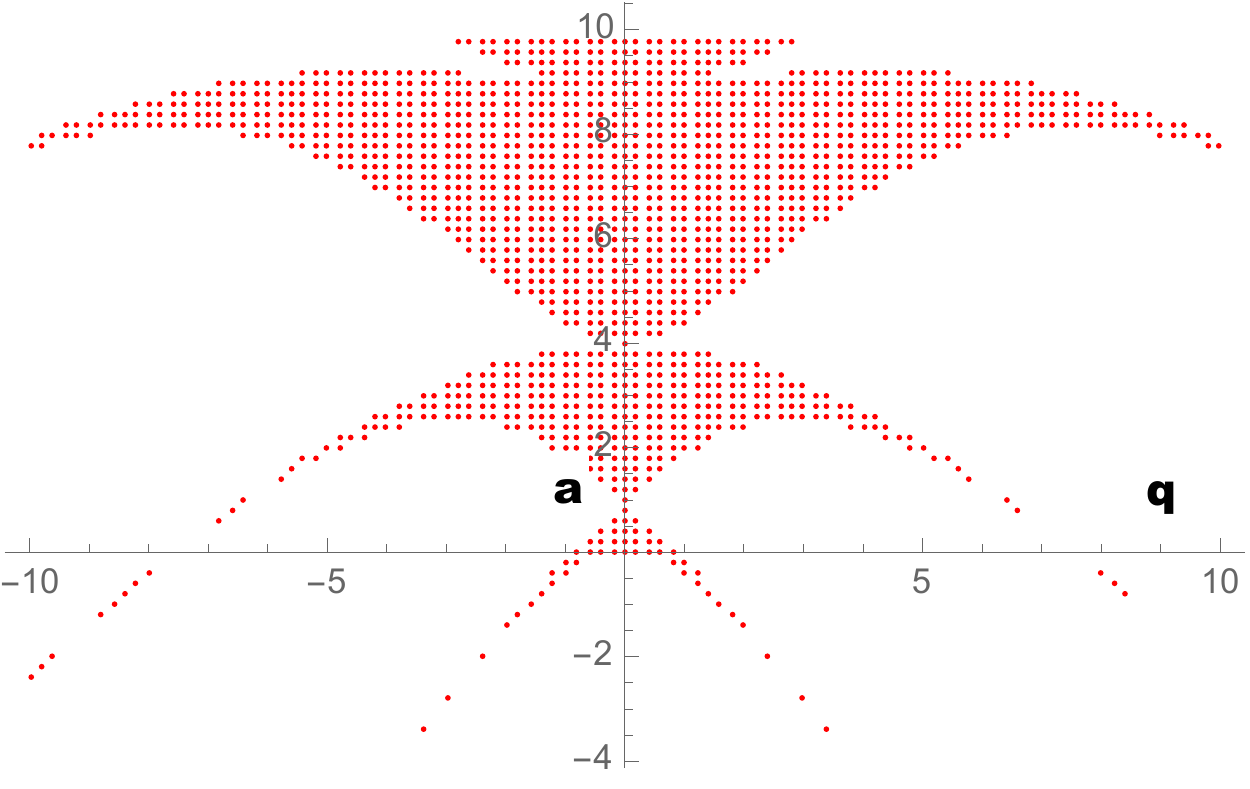}
\caption[Optional caption for list of figures]{\small Full stability region of a Paul trap in the presence of CSL. The dotted regions represent the stable regions which appear in form of bands.} 
\end{figure}

Next we choose a pair of points ($a,q$) very close to the boundary of the stability curve of the Mathieu equation, and by demanding that this point remains stable even with CSL, we put bounds on $r_c$ and $\lambda_{CSL}$.

\newpage

\begin{figure}[H]
\centering
  \includegraphics[width=9cm,height=6cm] {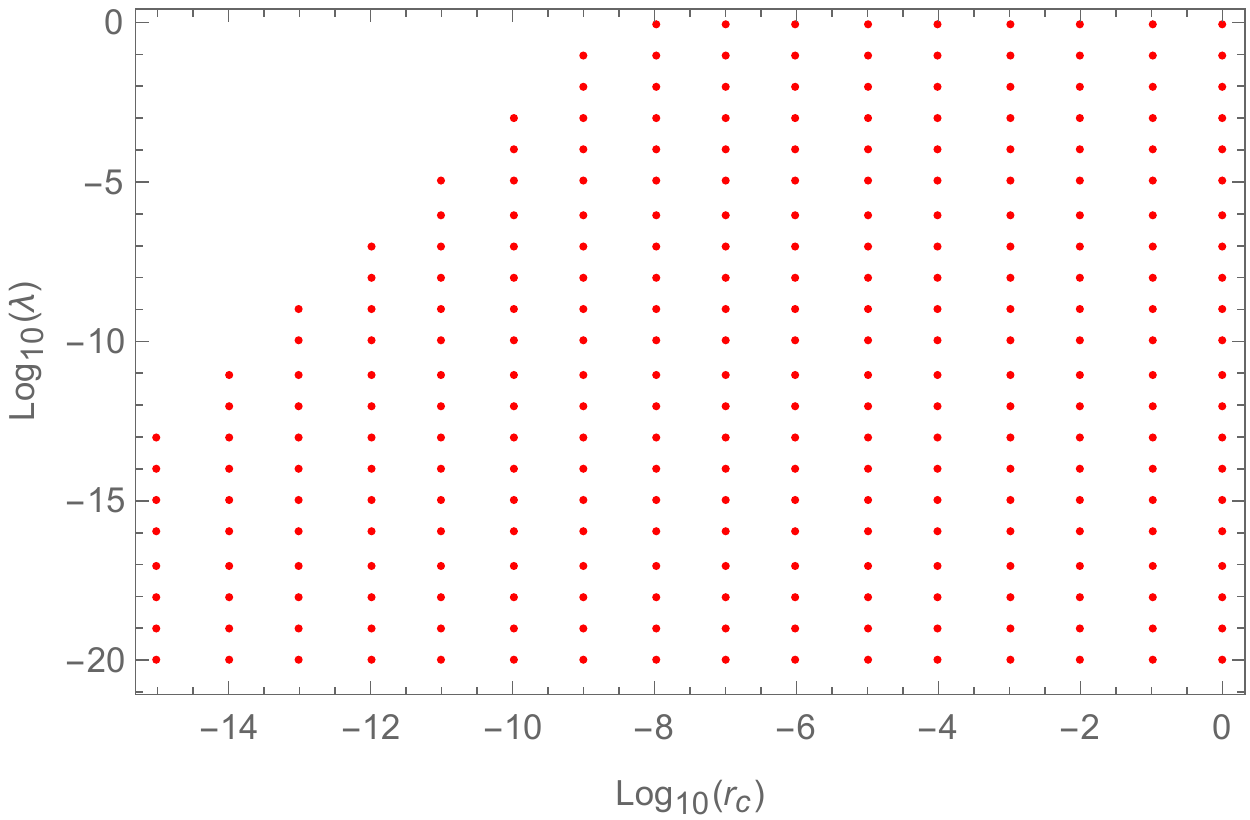}
\caption[Optional caption for list of figures]{$\lambda-r_c$ exclusion plot from the stability of charged particles in a Paul trap. The dotted region shows the allowed parameter space. The GRW and Adler parameter values ($\lambda_{GRW} = 10^{-17} s^{-1}$ and $\lambda_{ADLER} = 10^{-8} s^{-1}$) lie well within the allowed region.} 
\label{bound}
\end{figure}

 Below we show the allowed region in the $r_c-\lambda_{CSL}$ parameter space by choosing 
$a=-0.000526947, q= 0.0326158$ which is a stable point for Mathieu equation but very close to the boundary. This can be achieved, for example, for a trap configuration with RF oscillator frequency $\sim 100$ MHz, a dc voltage of the order of $\sim 70$ V, ac zero-to-peak voltage amplitude $\sim 8$ kV (sign of the voltage applied depends upon $x$ or $y$ direction) for a trapped ion with a mass-to-charge ratio of $\sim 100$ kg/Coulomb. These parameters are typically used in Paul trap experiments \cite{parameter1,parameter2}. Figure \ref{bound} shows the corresponding plot of $Log_{10} \lambda$ vs $Log_{10} r_c$. The dotted region represents the parameter space allowed by the stability analysis (i.e. they do not alter the stability of the above mentioned point). The white region represents unstable region. Although the bounds obtained are similar in nature to the bounds obtained from X-ray emission from an isolated Germanium slab in \cite{xray}, we find that the bounds obtained are rather weak and both $\lambda_{GRW}$ and $\lambda_{ADLER}$ are well inside the allowed region.

In an actual experiment involving Paul traps, there will be other factors present inevitably during the experiment that can produce significant amount of noise. In particular, one has to deal with the noises coming from fluctuations in the electromagnetic field, collision of the charged particle with ambient gas molecules, mechanical vibrations, production of eddy currents etc. very carefully to eliminate any disturbances that can jeopardize the stability of the trapped body. This has been discussed and taken into account in \cite{Bedingham}. In principle, with proper experimental parameters, such noises can be minimized and can be made subdominant to CSL effect, as has been shown in ref. \cite{Bedingham}.

\begin{figure}[H]
\centering
  \includegraphics[width=9cm,height=8cm] {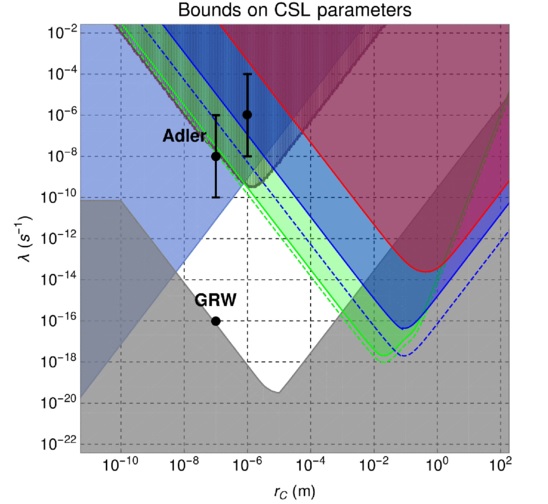}
\caption[Optional caption for list of figures]{Constraints on CSL parameters 
$\lambda_{CSL}$ and $r_c$ coming from various experiments. The plot has been taken from ref. \cite{CSLbound1}} 
\label{bassi}
\end{figure}

\section{Discussions}
Recently, with growing interests in testing collapse models, the field has seen a surge of interesting experimental ideas which can possibly give stringent bounds on CSL model parameters. A number of experiments have already ruled out a large portion of the parameter space. Recently in \cite{CSLbound1,CSLbound2}, the authors have put bounds on the CSL parameters coming from LIGO, LISA and AURIGA. If we also consider bounds on parameters based on purely  theoretical arguments, then the allowed parameter space becomes even smaller and $\lambda_{GRW}$ lies just at the boundary of the upper bounds coming from theoretical arguments \cite{thbound}. Figure \ref{bassi} shows various bounds on CSL parameters coming from different experiments. In the plot, the shaded regions of blue, green, and red lines are the exclusion regions as coming from LIGO, LISA Pathfinder, and AURIGA. Blue and green dashed lines represent upper bounds from foreseen improved sensitivity of LIGO and LISA Pathfinder, respectively. Purple line (towards the top) shows the upper bound from ultracold cantilever experiments \cite{cantilever}. Light blue region is excluded from x-ray experiments \cite{xray}. Gray region near the bottom is the exclusion zone based on theoretical arguments \cite{thbound}. The GRW and Adler values and ranges are indicated in black.  It would be interesting to check if $\lambda_{GRW}$ can be ruled out by experiments, and in that case we would have made the bounds even more stringent or possibly would invalidate the theory as a whole. At this juncture, it thus becomes very important to take care of each and every minute aspect in a proposed experiment, for example, various sources of noise, experimental setup, particle dynamics, careful and precise detection techniques etc. Any of these factors could affect the final outcome and lead to an erroneous conclusion. In this article, we discuss one of these factors that is effect of CSL dynamics on the stability of charged particles inside a Paul trap. Paul traps are used in such experiments to confine the particle in a small region so that the heating effect or the random displacement due to CSL can be observed. It is thus very important to make sure that the particle remains confined inside the stable region even when CSL noise force tries to displace it in random directions. Our study deals with the backreaction of CSL motion on the particle trajectory. The results show that for $\lambda_{GRW}$ parameters, there is no significant change in the stability zone. However, the points near the stable zone boundary can become unstable for certain values of the CSL parameter and hence extreme care should be taken while working in such a region. Ideally the trapping parameters should be chosen such that it lies well inside the stability zone, as for such cases, the CSL random motion will be ineffective to drive the particle out of its stable trajectory. \\

We have also plotted an exclusion plot for CSL parameters based on the Paul trap stability criteria. The bounds obtained are very similar to the constraints coming from X-ray experiments \cite{xray}, as can be seen in the plot above, but rather weaker. Future experiments are expected to produce stronger bounds and pushing the allowed region in the CSL parameter space further down before the theory can be ruled out with confirmation.

\section*{Acknowledgements}
We would like to thank Hendrik Ulbricht, Jerome Martin and Tejinder P. Singh for various helpful discussions and the organizers of the FPQP 2016 meeting held at ICTS, Bangalore (India) where the idea was conceived.

\end{document}